\documentclass[authoryear,12pt]{elsarticle}

 \journal{Foundations of Physics}

\usepackage{url,natbib}

\usepackage{amsmath, amssymb, amsfonts, amsthm, fouriernc}

\usepackage[usenames,dvipsnames]{color}

\newcommand{\Avec}{\mathbf{A}}
\newcommand{\Pvec}{\mathbf{P}}
\newcommand{\Vvec}{\mathbf{V}}
\newcommand{\dvec}{\mathbf{d}}
\newcommand{\evec}{\mathbf{e}}

\newcommand{\kvec}{\mathbf{k}}
\newcommand{\nvec}{\mathbf{n}}
\newcommand{\rvec}{\mathbf{r}}
\newcommand{\uvec}{\mathbf{u}}
\newcommand{\vvec}{\mathbf{v}}
\newcommand{\wvec}{\mathbf{w}}
\newcommand{\xvec}{\mathbf{x}}
\newcommand{\yvec}{\mathbf{y}}
\newcommand{\zvec}{\mathbf{z}}
\newcommand{\svec}{\mathbf{s}}
\newcommand{\Evec}{\mathbf{E}}
\newcommand{\Bvec}{\mathbf{B}}

\newcommand{\del}{\partial}

\DeclareMathOperator{\diag}{{\rm diag}}

\numberwithin{equation}{section}

\begin{document}

\begin{frontmatter}



\title{Free Electron Paths from Dirac's Wave Equation Elucidating Zitterbewegung and Spin}

\author[label1]{James L. Beck}
\address[label1]{California Institute of Technology, Mail Code 9-94, Pasadena. CA 91125, USA 
\newline  E-mail: jimbeck@caltech.edu}


\begin{abstract}
The worldline of a free electron is revealed by applying Dirac's velocity operator to its Dirac wave function whose space-time arguments are expressed in a proper time by a Lorentz transformation. 
This motion can be decomposed into two parts: the electron's global motion of its inertia (or spin) center and an inherent local periodic motion about this point that produces the electron's spin and 
has the zitterbewegung frequency found by Schr\"{o}dinger in his operator analysis of Dirac's wave equation. 
This zitter motion corresponds to the so-called polarization and magnetization currents in Gordon's decomposition of Dirac's current. 
In an inertial "rest"-frame fixed at the inertia center, Dirac's wave function for a free electron with its spin in a specified direction implies that the zitter motion is a perpetual circular motion 
about the inertia center in a plane orthogonal to this spin direction with a radius one half of the Compton radius and moving at the speed of light. 
The electron continuously accelerates about the spin center without any external force because the inertia is effective at the spin center, 
rather than at its charge center where the electron interacts with the electro-magnetic field. 
This analysis confirms the nature of zitterbewegung directly from Dirac's wave equation, agreeing with the conclusions of Barut and Zanghi, Beck, Hestenes, Rivas and Salesi 
from their classical Dirac particle models of the electron. Furthermore, these five classical models are equivalent and express the same free electron dynamics as Dirac's equation.
\end{abstract}
%

\end{frontmatter}


\section{Introduction}

The Dirac relativistic wave equation ({\color{Green}Dirac (1928)}) is arguably the best  particle model available for the dynamics of a single electron in the absence of particle creation and annihilation. 
It correctly predicts that the electron has spin with magnitude $\hbar / 2$ and it has a magnetic moment with $g=2$ that is very close to the experimentally observed value, 
although it does not seem to reveal the physics behind these values. It also accurately predicts the electro-magnetic spectrum for the hydrogen atom ({\color{Green}Dirac (1928)}). 
However, when Schr\"{o}dinger performed a Heisenberg operator analysis of Dirac's equation for a free electron, he found that it exhibits a component of motion 
with an ultra-high frequency $\omega_0 = 2 m c^2 / \hbar \sim 1.55 \times 10^{21}$s$^{-1}$ that he called \emph{zitterbewegung} (trembling motion) ({\color{Green}Schr\"{o}dinger (1930)}). 
Various explanations have been proposed over the last century for the source of this puzzling phenomenon but none have been widely accepted. 
A difficulty associated with it is that the electron appears to move at the speed of light $c$, even though it has a non-zero mass $m$; in fact, Dirac's velocity operators 
that appear in his equation have eigenvalues of magnitude $c$.  In this work, we will show that, contrary to common wisdom, Dirac's equation can reveal the physics behind these puzzling features. 
This analysis gives a resolution of several mysteries; for example, what is Schr\"{o}dinger's zitterbewegung that appears to have the electron moving at the speed of light; 
what is spin if it is not a rotation of the electron itself; and why does the electron, an apparent point charge, have dipoles? 

Dirac's equation appears to be quite distinct from classical mechanics models of the electron's dynamics since, instead of classical dynamic variables, 
it involves a spinor wave function and velocity and momentum operators that are not proportional through a scaling by its mass $m$. 
Furthermore, the wave function solutions are fields over space-time and do not seem to be associated with electron worldlines, again in contrast to classical relativistic mechanics. 
Nevertheless, various classical models of a charged particle with spin have been proposed for the electron that seem to bear a very close relationship with Dirac's model 
(e.g. {\color{Green}Barut and Zanghi (1984)}, {\color{Green}Beck (2023)}, {\color{Green}Hestenes (2010)}, {\color{Green}Rivas (2003)} and {\color{Green}Salesi (2002)}). 
These classical Dirac particle models simultaneously explain zitterbewegung and spin of an electron as an inherent circular motion about the inertia center at the speed of light $c$ with circular frequency equal to 
Schr\"{o}dinger's frequency $\omega_0$, implying a radius of $r_0 = c / \omega_0 = \hbar / (2mc) \sim 1.93 \times 10^{-13}$m related to the Compton wavelength and an angular momentum of $mcr_0 = \hbar / 2$. 
The center of this spin motion is the inertia center, which moves subluminally (in a straight line for a free electron), producing an overall helical spatial motion. 
Furthermore, these models give an explanation of the inherent magnetic moment of an electron as due to its effective loop current of the circulating electron. 

Although these classically-based models have been available for a long time, they have not had much impact on mainstream physics where Schr\"{o}-dinger's zitterbewegung, 
if it is mentioned at all, is usually described as "the interference between positive and negative energy solutions of Dirac's equation", 
which is based on the mathematics that produces the zitter motion but which lacks a physical picture of what is behind it. 
Actually, more than 70 years ago, Huang did provide a physical explanation of zitterbewegung based on analyzing a wave function expressed as 
a superposition of ``positive energy'' and ``negative energy'' solutions of Dirac's wave equation for a free electron ({\color{Green}Huang (1952)}). 
His analysis agrees with the predicted behavior of the electron given by the afore-mentioned classical Dirac particle models. 
Bohm and Hiley give a similar interpretation of Pauli's wave equation solution for the free electron in Chapter 10.4 of their book ({\color{Green}Bohm and Hiley (1993)}). 
They conclude that the spin angular momentum and magnetic moment of an electron come from a circulation of the electron as a point particle. 
These earlier studies have also had limited impact on the understanding of spin and zitterbewegung in mainstream physics.

In this work, we show that, despite the widespread belief that Dirac's equation does not reveal electron paths, they are hiding in plain sight. 
We derive the electron's worldline from the Dirac wave function for a free electron, that is, its space-time path as a function of a proper time, 
and show that the nature of zitterbewegung and spin are exactly as predicted by the afore-mentioned classical Dirac particle models.
In fact, we show that the model in {\color{Green}Barut and Zanghi (1984)}, which is based on a Lagrangian function involving both ``classical'' spinors and classical dynamic variables, 
is equivalent to Dirac's equation in the case of a free electron, that is, every Barut-Zanghi spinor solution is a Dirac spinor solution, and conversely. 

It has already been shown in {\color{Green}Beck (2023)} that there is an equivalence between the equations of motion of Beck's neoclassical relativistic mechanics model 
of the electron and the equations of motion in classical dynamic variables that are derived by applying Dirac's velocity and spin operators to the Barut-Zanghi spinor. 
Furthermore, as noted in {\color{Green}Rodriguez et al. (1993)}, the Barut and Zanghi model is equivalent to Hestenes' light-like zitter model in {\color{Green}Hestenes (1993, 2010)}, 
which also uses spinors with classical dynamical variables to express the dynamics of an electron. 
Beck's neoclassical model has the same Lagrangian function, given later, as the classical Dirac model in {\color{Green}Salesi (2002)} when minimal coupling is added to the latter. 
It also gives a covariant form of the equations of motion for the relativistic spinning particle model of Rivas 
that he derived from the Frenet-Serret equations for a curve in three-dimensional space ({\color{Green}Rivas (2003)}).
Therefore, all these classical Dirac particle models are equivalent to each other in their description of the dynamics of the spinning electron, 
and for the free electron, they all agree in this sense with Dirac's wave function model. 
Furthermore, the  electron's motion in the classical Dirac models can be decomposed into a global part, which is the motion of the inertia (or spin) center, 
and a local spin part that is Schr\"{o}dinger's zitterbewegung, and we show that for the electron's velocity, this corresponds to the Gordon decomposition of the Dirac current.

\section{Solution of Covariant Dirac Equation for a Free Electron}

Consider a \emph{free} electron of mass $m$ and charge $q = -e$ (so $e>0$ is the unit electronic charge), although the charge plays no role for a free electron. 
Actually, it seems that the theory presented here can be applied to any lepton with the appropriate choice of mass $m$ and charge $q$. 

Let $x^{\mu}$ and $\pi^{\mu}$ ($\mu = 0,1,2,3$) be the contravariant components of an electron's 4-vector position $x = (ct,\xvec)$ and 
its constant 4-vector momentum $\pi = (E/c, \Pvec)$, relative to an observer inertial reference frame $X_o$ with its origin at some point $O$. 
Throughout the paper, bold letters are 3-vectors with ${\bf a} \cdot {\bf b}$ and ${\bf a} \times {\bf b}$ denoting the scalar and vector products and 
Einstein's summation convention for repeated Greek letter indices is used. The Minkowski metric tensor is $G = \diag(1,-1,-1,-1)$ with components $g_{\mu\nu} = g^{\mu\nu}$.

The wave function $\psi(x) \in {\mathbb{C}}^4$ for the free electron satisfies Dirac's wave equation written in covariant form ({\color{Green}Bjorken and Drell (1964)}) as 
\begin{equation} \label{Eq_1_1}
	\widetilde{H} \psi(x) \equiv \hat{u}^{\mu} \hat{\pi}_{\mu} \psi(x) = mc^2 \psi(x) 
\end{equation}
\noindent where Dirac's 4-velocity operator $\hat{u}^{\mu} = c \gamma^{\mu} = c\beta \alpha^{\mu}$ wth the usual Dirac matrices ($\alpha^0 = I_4$, the identity matrix), 
and his 4-momentum operator $\hat{\pi}_{\mu} = i \hbar \del_{\mu} = i \hbar \frac{\del} {\del x^{\mu}}$ (since the em potential is zero, the kinematic and canonical momenta are equal). 
We can write the general solution of this equation as
\begin{equation} \label{Eq_1_2}
	\psi(x) = \left[ \cos \theta(x) I_4 - \frac{i}{mc^2} \sin \theta(x) \hat{H} \right] A
\end{equation}
\noindent where $A \in {\mathbb{C}}^4$ is an arbitrary constant so far. The phase 
\begin{equation} \label{Eq_1_3}
	\theta(x) \equiv x^{\mu} \pi_{\mu} / \hbar = (Et - \Pvec \cdot \xvec) / \hbar = \omega t - \kvec \cdot \xvec 
\end{equation} 
\noindent gives the only dependence of $\psi$ on $x$ and the Barut-Zanghi Hamiltonian operator $\hat{H} =  \hat{u}^{\mu} \pi_{\mu} \in {\mathbb{C}}^4$ satisfies the identity 
\begin{equation} \label{Eq_1_4}
	\hat{H}^2 = c^2 \pi_{\mu} \pi_{\nu} \gamma^{\mu} \gamma^{\nu} = \frac{1} {2} c^2 \pi_{\mu} \pi_{\nu} \left( \gamma^{\nu} \gamma^{\mu} + \gamma^{\mu} \gamma^{\nu} \right) 
	= c^2 \pi_{\mu} \pi_{\nu} g^{\nu\mu} I_4 = (mc^2)^2  I_4
\end{equation}
\noindent where the final equality comes from the usual energy-momentum equation 
\begin{equation} \label{Eq_1_5}
	\pi_{\mu} \pi^{\mu} = (E/c)^2 - \Pvec^2 = (mc)^2 
\end{equation}
\noindent for a free electron. This energy equation can be derived from the invariance of the 4-vector inner product under a Lorentz transformation by choosing another 
inertial reference frame $X_c$ for which $\Pvec = \mathbf{0}$ and $E=mc^2$. We will call $X_c$ \emph{a rest-frame} of the electron because it has zero momentum relative to $X_c$. 
There are an infinite number of such rest-frames, of course, depending on the choice of the origin $C$ for $X_c$, but we will choose a special point for $C$ later to give 
\emph{the} rest-frame of the electron. (For convenience, we take the directions of the spatial axes of $X_o$ and $X_c$ as aligned).

Notice that the form of the phase in Eq.~\eqref{Eq_1_3} endows $\psi(x)$ with an apparent plane-wave characteristic with the well-known de Broglie frequency-energy 
and wavenumber-momentum relations, $E = \hbar \omega$ and $\Pvec = \hbar \kvec$. 
However, this "wave'' has an apparent wave propagation speed of $E/| \Pvec | > c$ when viewed by an observer fixed with respect to 
the reference frame $X_o$, and vanishes when viewed from the rest-frame $X_c$, so it is not a physical wave. 

Using the expression for $\psi(x)$, we can derive for $\mu = 0,1,2,3$, 
\begin{equation} \label{Eq_1_6}
	\bar{\psi}(x) \hat{\pi}_\mu \psi(x) = \frac{\pi_\mu} {mc^2} \bar{A} \hat{H} A = \pi_\mu
\end{equation}
\noindent where we have the usual definition $\bar{\psi} = \psi^{*} \gamma^0$ with $\psi^{*}$ the Hermitian (conjugate) transpose of $\psi$, 
and we have normalized $A$ so that $\bar{A} \hat{H} A = mc^2$. In deriving this equation, we used the result that $\hat{H}^{*} = \gamma^{0} \hat{H} \gamma^{0}$. 
Notice also that $\psi(0) = A$.

In a similar way, we can derive Dirac's 4-velocity for $\mu = 0,1,2,3$, 
\begin{align}  \nonumber
	\widetilde{u}^{\mu}(x) & \equiv \bar{\psi}(x) \hat{u}^{\mu} \psi(x) \\ 
	& = \bar{A} \left[ \cos^2 \theta(x) \hat{u}^{\mu} + \frac{i}{2mc^2} \sin 2 \theta(x) \left[ \hat{H}, \hat{u}^{\mu} \right] 
		 +  \frac{1}{(mc^2)^2} \sin^2 \theta(x)~\hat{H} \hat{u}^{\mu} \hat{H} \right] A  \label{Eq_1_7} 
\end{align}
\noindent where we introduce the commutator $\left[ \hat{H}, \hat{u}^{\mu} \right] =  \hat{H}\hat{u}^{\mu} - \hat{u}^{\mu}\hat{H}$.
Furthermore, using the standard identity for the Dirac matrices twice 
\begin{align}  \nonumber 
	\hat{H} \hat{u}^\mu \hat{H} & =  c^3 \pi_\nu \pi_\sigma \gamma^\nu \gamma^\mu \gamma^\sigma  \\ \nonumber
	& = - c^3 \pi_\nu \pi_\sigma \gamma^\nu \gamma^\sigma \gamma^\mu + 2 c^3 \pi_\nu \pi_\sigma g^{\mu\sigma} \gamma^\nu  \\ \nonumber
	& = - \frac{c^3}{2} \pi_\nu \pi_\sigma \left( \gamma^\nu \gamma^\sigma +  \gamma^\sigma \gamma^\nu \right)  \gamma^\mu + 2c^2 \pi^\mu \hat{H}  \\ 
	& = - c^3 \pi_\nu \pi^\nu  \gamma^\mu + 2c^2 \pi^\mu \hat{H}  =  - \left( mc^2 \right)^2  \hat{u}^\mu + 2c^2 \pi^\mu  \hat{H}  \label{Eq_1_8} 
\end{align}
\noindent Introduce the operator $\hat{a}^{\mu} = i \hbar^{-1} \left[ \hat{H}, \hat{u}^{\mu} \right]$, define $ \widetilde{a}^{\mu} = \bar{A} \hat{a}^{\mu} A$ 
and note that $\widetilde{u}^{\mu}(0) = \bar{A} \hat{u}^{\mu} A$ since $\psi(0) = A$, then substituting these results into Eq.~\eqref{Eq_1_7}
\begin{align}  \nonumber
	\widetilde{u}^{\mu}(x) & = \frac{1}{m} \pi^\mu + \left(  \widetilde{u}^\mu(0) - \frac{1}{m} \pi^\mu \right) \cos 2\theta(x) + \frac{1}{\omega_0}  \widetilde{a}^\mu \sin 2\theta(x) \\ 
	& = v^\mu + \widetilde{w}^\mu(x) \label{Eq_1_9}
\end{align}
\noindent using the normalization of $A$ given above and Schr\"{o}dinger's zitterbewegung frequency $\omega_0 = 2mc^2/ \hbar$. 
This equation can be interpreted as the velocity of the electron when it is at the space-time point $x = (ct,\xvec)$. 
This interpretation is not novel but it is not the standard interpretation which regards $\widetilde{u}^{\mu}(x)$ as the expected value of the velocity if the electron is located at $x$. 

It will become evident in the next subsection that the contribution $v^\mu = \frac{1}{m} \pi^\mu$ to $\widetilde{u}^\mu$ is the global velocity of the electron's inertia center 
while the contribution $\widetilde{w}^\mu$ is the velocity of the zitter motion, which is a local circular motion about the inertia center.  
Notice also that the time component $\widetilde{u}^0(x) = \psi(x)^* \psi(x) = E/(mc) + \widetilde{w}^0(x)$ is not the probability density function $\rho(x)$ for the electron's location, as usually postulated 
for Dirac's theory, since for a free electron, $\rho(x)$ is usually considered to be constant (Section 2.4 has more about the uncertainty in the electron's position and velocity).

As noted in {\color{Green}Salesi (2002)} but not proved there, Eq.~\eqref{Eq_1_9} corresponds to the covariant form of the Gordon decomposition of the Dirac current 
(dropping the charge $q$ as a multiplier) 
\begin{equation} \label{Eq_1_9b}
	\widetilde{u}^\mu (x) = \frac{1}{m} Re[\bar{\psi}(x) \hat{\pi}^\mu \psi(x)] - \frac{1}{m} \del_\nu \widetilde{S}^{\mu\nu} (x)
\end{equation} 
\noindent where the 4-tensor for spin $\widetilde{S}^{\mu\nu} (x) = \bar{\psi}(x) \hat{S}^{\mu\nu} \psi(x)$ and Dirac`s spin tensor operator is defined by 
$\hat{S}^{\mu\nu} = - \frac{i \hbar}{4} \left[ \gamma^{\mu}\gamma^{\nu} - \gamma^{\nu}\gamma^{\mu} \right]$. 
In the free electron case, $\psi(x)$ is given by Eq.~\eqref{Eq_1_2} and so by Eq.~\eqref{Eq_1_6}, the first term gives $\frac{1}{m} \pi^\mu = v^\mu$.
From Eq.~\eqref{Eq_1_9}, we can therefore infer that 
\begin{equation} \label{Eq_1_9c}
	- \frac{1}{m} \del_\nu \widetilde{S}^{\mu\nu} (x) = \widetilde{w}^\mu(x) \equiv \left(  \widetilde{u}^\mu(0) - \frac{1}{m} \pi^\mu \right) \cos 2\theta(x) + \frac{1}{\omega_0}  \widetilde{a}^\mu \sin 2\theta(x) 
\end{equation} 
\noindent We will show this result explicitly in Section 3.2 where an expression for $\widetilde{S}^{\mu\nu} (x)$ is derived. 

Because of the structure of the 4-tensor for spin, the spatial vector part of $\frac{q}{m} \del_\nu \widetilde{S}^{\mu\nu} (x)$ can be expressed as the sum 
of polarization and magnetization currents (e.g. {\color{Green}Baym (1981)}), as shown later in Section 3.2. 
Therefore, these currents are a consequence of the zitter motion of the electron.   
Furthermore, the Gordon decomposition in Eq.~\eqref{Eq_1_9b} corresponds to expressing the total current $q \widetilde{u}^\mu(x)$ as the sum of a point charge current $\frac{q}{m} \pi^\mu (x)$ 
and a dipole current $\del_\nu \widetilde{P}^{\mu\nu} (x)$ where the dipole moment tensor $\widetilde{P}^{\mu\nu} (x) = - \frac{q}{m} \widetilde{S}^{\mu\nu} (x)$ (see Eq. (1)-(4) in {\color{Green}Peletminskii and Peletminskii (2005)}). 
This implies that minimal coupling to an em-field (electro-magnetic field) with potential $A^\mu (x)$ through $q \widetilde{u}^\mu A_\mu$ also gives a dipole coupling.

\subsection{Electron's Wave Function and Space-time Motion in Proper Time}

Because of the invariance of the 4-vector inner product under a Lorentz transformation between the inertial reference frames $X_o$ and $X_c$, 
the phase in $\psi(x)$ can also be expressed as
\begin{equation} \label{Eq_1_10}
	\theta(x) \equiv x^{\mu} \pi_{\mu} / \hbar = \frac{\omega_0 }{2} \tau(x)
\end{equation} 
\noindent where the space-time coordinates of the electron relative to $X_c$ are $x_c = (c\tau, \rvec)$ and, by the definition of rest-frame $X_c$, its 4-momentum is $\pi_c = (mc, \mathbf{0})$. 
We call $\tau(x)$ the \emph{proper time} of the electron when it is located at $x$ relative to $X_o$. It is the time of a clock fixed at C, the origin of $X_c$.

Since the wave function $\psi(x)$ in Eq.~\eqref{Eq_1_2} can be expressed purely in terms of $\tau(x)$, we can view it as a function of $\tau$ and write it as $\phi(\tau)$ so that 
\begin{equation} \label{Eq_1_15}
	\phi(\tau) = \left[ \cos (\omega_1 \tau) I_4 - \frac{i}{mc^2} \sin (\omega_1 \tau) \hat{H} \right] A
\end{equation}
\noindent where $\omega_1 = \omega_0 /2$, showing that $\phi(0) = A$. 
In fact, this is just the wave function $\psi_c(\tau)$ expressed relative to $X_c$ boosted to give the wave function $\psi(x) = \phi(\tau(x))$ relative to $X_o$. 
Deboosting from $X_o$ to $X_c$ transforms $x$ to $\tau$ because Eq.~\eqref{Eq_1_10} corresponds to the time part of the Lorentz transformation of $x = (ct,\xvec)$ to 
$x_c = (c\tau, \rvec)$, as shown in {\color{Green}Beck (2023)} in Section 2.2. There is no dependence of the rest-frame wave function on $\rvec$ because $\Pvec_c = \mathbf{0}$.

Similarly, instead of viewing $\widetilde{u}^\mu(x)$ as a field over space-time coordinates $x$, we can view it as a function of the electron's proper time, so Eq.~\eqref{Eq_1_9} can be re-written as 
\begin{align}  \nonumber
	u^\mu(\tau) \equiv \bar{\phi}(\tau)~\hat{u}^\mu \phi(\tau) & = \frac{1}{m}\pi^\mu + \left( \widetilde{u}^\mu(0) - \frac{1}{m} \pi^\mu \right) \cos \omega_0\tau + \frac{1}{\omega_0} \widetilde{a}^\mu \sin \omega_0\tau \\ 
	& = v^\mu + w^\mu(\tau) \label{Eq_1_11}
\end{align}
\noindent The 4-vector proper velocity $u^\mu(\tau)$ is therefore a superposition of the constant global velocity $v^\mu = \pi^\mu/m$ and 
an oscillatory velocity $w^\mu(\tau)$ with angular frequency $\omega_0$ that is a manifestation of Schr\"{o}dinger's zitterbewegung. 
The same expression for the proper velocity was derived in {\color{Green}Barut and Zanghi (1984)}, {\color{Green}Salesi (2002)} and {\color{Green}Beck (2023)} 
for their respective classical models of the Dirac electron. 
Notice that with the new notation, the previous velocity fields $\widetilde{u}^\mu(x)$ and $\widetilde{w}^\mu(x)$ can be written as $u^\mu(\tau(x))$ and $w^\mu(\tau(x))$; and similarly, $\psi(x)$ is $\phi(\tau(x))$. 

Denote the derivative with respect to proper time with an overhead dot.
Notice that $\dot{t}(\tau) = u^0(\tau) /c = \bar{\phi}(\tau)~\hat{u}^0 \phi(\tau) /c = \phi(\tau)^* \phi(\tau) > 0$, so observer time $t(\tau)$ increases monotonically with proper time $\tau$, 
although it has an ultra-high frequency oscillation about a linear trend that arises from integrating the constant rate $\pi^0 /(mc) = E/(mc^2)$ in Eq.~\eqref{Eq_1_11}.
The electron's 4-vector acceleration is
\begin{equation} \label{Eq_1_12}
	\dot{u}^\mu (\tau) = - \omega_0 \left(  \widetilde{u}^\mu(0) - \frac{1}{m} \pi^\mu \right) \sin \omega_0 \tau +  \widetilde{a}^\mu \cos \omega_0 \tau
\end{equation} 
Setting $\tau=0$ in the last two equations, we see that $\widetilde{u}^\mu(0) = u^{\mu}(0)$ and $\widetilde{a}^\mu = \dot{u}^\mu (0)$.

We now interpret the 4-velocity $u^{\mu}(\tau)$ as $\dot{x}^{\mu}(\tau)$, the rate of change of $x^{\mu}(\tau)$, the position of the electron at proper time $\tau$. Integrating Eq.~\eqref{Eq_1_11}
\begin{align}  \nonumber
	x^\mu (\tau) & = \left[ \frac{1}{m} \pi^\mu \tau + y^\mu(0) \right] + \left[ \frac{1}{\omega_0} \left( u^\mu(0) - \frac{1}{m} \pi^\mu \right) \sin \omega_0 \tau -  \frac{1}{\omega_0^2} \dot{u}^\mu (0) \cos \omega_0 \tau \right] \\
 	& = y^\mu (\tau) + z^\mu  (\tau)  \label{Eq_1_13}
\end{align}
\noindent where the constant $y^\mu(0) = x^\mu(0) + \dot{u}^\mu(0) / \omega_0^2 = x^\mu(0) - z^\mu(0)$. 
Actually, this interpretation is consistent with the Heisenberg operator for rate of change with respect to proper time that was introduced in {\color{Green}Beck (1942)} 
and also derived rigorously in {\color{Green}Barut and Thacker (1985)}. Here, the Beck derivative operator gives 
\begin{equation} \label{Eq_1_11b}
	 \hat{\dot{x}}^\mu = - \frac{i}{\hbar}  \left[ \widetilde{H}, x^\mu  \right] = \hat{u}^\mu
\end{equation}
\noindent and so $\dot{x}^\mu(\tau) = \bar{\phi}(\tau)~\hat{\dot{x}}^\mu \phi(\tau) = \bar{\phi}(\tau)~\hat{u}^\mu \phi(\tau) = u^\mu(\tau)$. 
We also see from its definition in Eq.~\eqref{Eq_1_13} that $z^\mu(\tau)$ satisfies the equation of motion $\ddot{z}^\mu = - \omega_0^2 z^\mu$. 

The electron's space-time path $x^{\mu}(\tau)$ can therefore be expressed as the sum of two parts, a \emph{global motion} 
$y^\mu (\tau) = \frac{1}{m} \pi^\mu \tau + y^\mu(0)$, and a \emph{local motion} $z^\mu  (\tau)$ consisting of the two periodic terms in Eq.~\eqref{Eq_1_13}. 
The worldline $y^\mu (\tau)$ of the inertia center and its velocity $\dot{y}^\mu (\tau) = \frac{1}{m} \pi^\mu = v^\mu$ are what is expected classically for a free electron 
with constant momentum $\pi^\mu = m v^\mu$. If we write $y = (ct_y, \yvec)$, then $\dot{y} = (c\gamma, \vvec) = \frac{1}{m} \pi$, so $E = \gamma m c^2$ and 
$\Pvec = m \vvec = \gamma m \Vvec$ where $\gamma = \dot{t}_y$ and $\vvec = \frac{d \yvec}{d \tau} = \gamma \frac{d \yvec}{d t_y} = \gamma \Vvec$. 
By substituting in Eq.~\eqref{Eq_1_5}, we get the usual expression 
\begin{equation} \label{Eq_1_11c}
	\gamma = \left[ 1 - \Vvec^2/c^2 \right]^{-1/2}
\end{equation}
 \noindent The contribution $z(\tau) = (ct_z, \zvec)$ to the \emph{total motion} $x(\tau)$ is a local motion of the electron about $\yvec(\tau)$ since it remains spatially bounded for all time. 
It is periodic with frequency $\omega_0$ and corresponds to Schr\"{o}dinger's zitterbewegung. Following Hestenes, we will call $z(\tau)$ the electron's \emph{zitter motion}, 
although, as shown below, it is the source of the electron's spin, so \emph{spin motion} would also be appropriate.

\subsection{An Aside on Normalization of Wave Functions}

The normalization of the wave function $\psi$ so that the constant spinor $A$ satisfies $\bar{A} \hat{H} A = mc^2$ is seen to be essential for the Dirac operators 
$\hat{u}^\mu$ and $\hat{\pi}_\mu$ to produce the proper velocity $u^\mu = \frac{dx^\mu}{d\tau}$ and momentum $\pi^\mu = m \frac{dy^\mu}{d\tau}$ from the wave function. 
It can be readily shown from Eq.~\eqref{Eq_1_2} that the chosen normalization of $A$ implies that the wave function satisfies $\bar{\psi}(x) \hat{H} \psi(x) = mc^2$ 
over any space-time region, or, equivalently, $\bar{\phi}(\tau) \hat{H} \phi(\tau) = mc^2$ for all proper time. 

If observer time is used for the rate of change, the velocity $U^\mu = \frac{dx^\mu}{dt}$ and momentum $\Pi^\mu = m \frac{dy^\mu}{dt}$ are invariant to a 
constant scaling of $\psi$. For example,
\begin{equation} \label{Eq_1_11d}
	U^\mu = \frac{dx^\mu} {dt} = \widetilde{u}^\mu / \widetilde{\dot{t}} =  \frac{\bar{\psi}(x) \hat{u}^\mu \psi(x)} {\psi^*(x) \psi(x)}
\end{equation}
\noindent because $\widetilde{\dot{t}}(x) = \widetilde{u}^0(x) / c = \bar{\psi}(x)~\hat{u}^0 \psi(x) /c = \psi(x)^* \psi(x)$. 
If $\psi(x)$ is normalized so that its Euclidean norm is unity, as sometimes done, then it forces $\widetilde{\dot{t}}(x) = 1$, which holds exactly only relative to the rest-frame $X_c$, 
and approximately in non-relativistic mechanics, but not in general.

\subsection{Rest-frame Zitter Motion of Electron}

We have not yet defined the origin $C$ of $X_c$ but it must have global velocity $\vvec_c = \mathbf{0}$ relative to $X_c$ in order to get $\Pvec_c = \mathbf{0}$, 
that is, the electron is globally at rest relative to $X_c$, which is why we called it a rest-frame of the electron. 
The 4-vector global velocity relative to $X_c$ is $\dot{y}_c = v_c = \pi_c / m = (E_c/(mc), \vvec_c) = (c, \mathbf{0})$, so its time component is not zero. 
Relative to $X_c$, we have from Eq.~\eqref{Eq_1_11c} that  $\dot{t}_y = \gamma = 1$ and so we can take $t_y = \tau$. 

We choose origin $C$ to be the point $\yvec(\tau)$ relative to $X_o$ so that it moves with the global motion of the electron relative to $X_o$. 
This means that $C$ is chosen as the point where the momentum of the electron is located ($\Pvec = m \vvec = m \dot{\yvec}$, not $m\uvec$). 
Clearly, the inertia of the electron is located at the point $C$ and not at the electron's spatial position $\xvec(\tau)$ where the charge is located, 
so $C$ is appropriately called the \emph{inertia center} (or center of mass) of the electron, although, as we will see, it is also appropriate to call it the \emph{spin center}. 
This physical separation of the inertia and charge centers of an electron is a characteristic of the afore-mentioned classical Dirac particle models 
(see {\color{Green}Rivas (2015)} for a discussion of this feature). 
Furthermore, since the free electron is perpetually accelerating (see Eq.~\eqref{Eq_1_12}), 
it is clear that the electron does not emit or absorb radiation (photons) unless its inertia center accelerates. 
This opens a new window into the nature of the electron's inertia and mass that could be further explored. 

Relative to $X_c$, the global motion of the electron is $\yvec_c (\tau) = \vvec_c \tau = \mathbf{0}$. Thus, the motion $\xvec_c(\tau) = \zvec_c(\tau)$ is the zitter part. 
From Eq.~\eqref{Eq_1_13}: 
\begin{equation} \label{Eq_1_14}
	\rvec(\tau) = \xvec_c(\tau) = \zvec_c(\tau) = \frac{1}{\omega_0} \dot{\rvec}(0) \sin \omega_0 \tau -  \frac{1}{\omega_0^2} \ddot{\rvec}(0) \cos \omega_0 \tau 
\end{equation}
\noindent implying that $\ddot{\rvec}(\tau) = - \omega_0^2 \rvec(\tau)$.

If we impose the condition $\dot{\rvec}(\tau) \cdot \dot{\rvec}(\tau) = c^2$, then we can conclude from Eq.~\eqref{Eq_1_14} that $\rvec(\tau) \cdot \rvec(\tau) = r_0^2$, a constant (see Appendix A). 
Thus, relative to $X_c$, the electron moves in a circle about $C$ with speed $c$. Clearly, its circular frequency is $\omega_0$ and so the radius $r_0 = c / \omega_0 = \hbar / (2mc)$. 
However, there is widespread belief that a particle with mass cannot move at the speed of light, so to address this issue, 
we next show directly from the free Dirac wave function that the zitter motion is luminal, although the observed global motion of the electron that carries its inertia is sub-luminal. 

We can re-write the wave function solution in Eq.~\eqref{Eq_1_15} in terms of complex exponentials as
\begin{equation} \label{Eq_1_16}
	\phi(\tau) = e^{-i \omega_1 \tau}  A^+ + e^{i \omega_1 \tau}  A^-
\end{equation}
\noindent where we define
\begin{equation} \label{Eq_1_17}
	A^+ = \frac{1}{2} \left(I_4 + \frac{1}{mc^2} \hat{H} \right) A, \hspace{0.2in} A^- = \frac{1}{2} \left(I_4 - \frac{1}{mc^2} \hat{H} \right) A
\end{equation}
\noindent The wave function solution in this form is said to be a superposition of positive energy and negative energy solutions, although this is misleading terminology
since the latter solutions do not really represent negative energy. Furthermore, they are essential to correctly localize the electron in its path (see related comments later in this subsection). 
Mistakenly dropping them because of the belief that they are unphysical, as some authors have done, is like ``throwing the baby out with the bathwater''.

Using the identity Eq.~\eqref{Eq_1_4}, we have $\hat{H} A^+ = mc^2 A^+$ and $\hat{H} A^- = - mc^2 A^-$, showing that $A^+$ and $A^-$ are eigenvectors of $\hat{H}$ 
that depend only on the global momentum $\pi$; they lie in the two-dimensional eigenspaces corresponding to eigenvalues $mc^2$ and $-mc^2$, respectively. 

To examine the nature of the zitter motion, consider the case where the observer's inertial reference frame $X_o$ coincides with $X_c$, the rest-frame fixed at $C$, 
then $\hat{H} = mc^2 \gamma^0$ and so the eigenvectors of $\hat{H}$ have the form:
\begin{equation} \label{Eq_1_18}
	A^+ = \left[ A_1, A_2, 0,0 \right]^T, \hspace{0.2in} A^- = \left[ 0,0, A_3, A_4 \right]^T
\end{equation}
\noindent We can therefore express the wave function solution in Eq.~\eqref{Eq_1_16} as
\begin{equation} \label{Eq_1_19}
	\phi(\tau)  = \left[ A_1 e^{-i \omega_1 \tau}, A_2 e^{-i \omega_1 \tau}, A_3 e^{i \omega_1 \tau}, A_4 e^{i \omega_1 \tau} \right]^T
\end{equation}
\noindent The complex amplitudes $A_j = \phi_j(0)$  $(j=1,...,4)$ must satisfy the previous energy normalization of $A$, so $mc^2 = \bar{A} \hat{H} A = mc^2 A^* A$, which implies: 
\begin{equation} \label{Eq_1_20}
	1 = A^* A = |A_1|^2 + |A_2|^2 + |A_3|^2 + |A_4|^2
\end{equation}
\noindent This is consistent with $\phi^*(\tau)  \phi(\tau) = \frac{1}{c} \bar{\phi}(\tau) \hat{u}^0 \phi(\tau) = \frac{1}{c} u^0(\tau) = \dot{t}(\tau) =  1$ 
because $t=\tau$ in the rest-frame at $C$. 
The next equation shows that for a spin state, $ |A_1|^2 + |A_2|^2 | = \frac{1}{2} = |A_3|^2 + |A_4|^2 $, so the positive and negative energy solutions are of equal amplitude. 

We can apply Dirac's 4-vector velocity operator and his spin tensor operator to the wave function $\phi(\tau)$ in Eq.~\eqref{Eq_1_19}. 
This analysis has already been done in Section 4.3 of {\color{Green}Beck (2023)} since the Barut-Zanghi spinor solution that he used is also a Dirac wave function. 
In particular, it is shown there that the complex coefficients $A_i$ corresponding to the state of an electron spinning about some direction 
defined by a unit vector $\nvec$, are given by:
\begin{align}  \nonumber
	A_1 & = \frac{1}{\sqrt{2}} \exp (-i \varphi/2) \cos (\theta/2), \hspace{0.1in} & A_2 = \frac{1}{\sqrt{2}} \exp (i \varphi/2) \sin (\theta/2) \\
	A_3 & = - \frac{1}{\sqrt{2}} \exp (-i \varphi/2) \sin (\theta/2), \hspace{0.1in} & A_4 = \frac{1}{\sqrt{2}} \exp (i \varphi/2) \cos (\theta/2)  \label{Eq_1_21}
\end{align}
\noindent where $\nvec = (n^1, n^2, n^3) = (\sin \theta \cos \varphi, \sin \theta \sin \varphi, \cos \theta) = n^j \evec_j$ (summation convention over $j=1,2,3$); 
$\{\evec_1, \evec_2, \evec_3 \}$ is an orthonormal triad of 3-vectors; and spherical coordinates are used with polar angle $\theta$ and the $x^3$-axis as the polar axis. 
These coefficients give the Dirac wave function $\phi_n(\tau)$ that is an eigenvector corresponding to the positive eigenvalue of the spin operator 
\begin{equation} \label{Eq_1_22}
	\hat{s}_n = n^j \hat{s}^j = \frac{\hbar}{2} \left[  \begin{array}{cc} 
		\sigma_n & 0 \\
		0 & \sigma_n 
\end{array} \right] 
\end{equation}
\noindent Here $\sigma_n = n^j \sigma^j$ where the $\sigma^j$'s are the three Pauli spin matrices, so 
\begin{equation} \label{Eq_1_23}
	\sigma_n = \left[  \begin{array}{cc} 
		\cos \theta & \exp (-i \varphi) \sin \theta \\
		\exp (i \varphi) \sin \theta & - \cos \theta 
\end{array} \right] 
\end{equation}

Expressions for the spatial velocity components $u^\mu = \bar{\phi} \hat{u}^\mu \phi$ and spin components $s^\mu = \bar{\phi} \hat{s}^\mu \phi$ for $\mu = 1,2,3$ are given 
in Appendix C of {\color{Green}Beck (2023)} for arbitrary coefficients $A_j$ in Eq.~\eqref{Eq_1_19}. Substituting the values given in Eq.~\eqref{Eq_1_21} 
into these expressions gives the velocity and spin components relative to $X_c$ for the spin state $\phi_n (\tau)$ (we suppress here the subscript $c$ on $u$ and $s$):
\begin{align}  \nonumber
	u^1 & = c \cos^2 \frac{\theta}{2} \cos(\omega_0 \tau + \varphi) - c \sin^2 \frac{\theta}{2} \cos(\omega_0 \tau - \varphi)  \\  \nonumber
		& = c \cos \theta \cos \varphi \cos (\omega_0 \tau) - c \sin \varphi \sin (\omega_0 \tau) \\  \nonumber
	u^2 & = c \cos^2 \frac{\theta}{2} \sin(\omega_0 \tau + \varphi) + c \sin^2 \frac{\theta}{2} \sin(\omega_0 \tau - \varphi)  \\  \nonumber
		& = c \cos \varphi \sin (\omega_0 \tau) + c \cos \theta \sin \varphi \cos (\omega_0 \tau) \\
	u^3 & =  - c \sin \theta  \cos (\omega_0 \tau)  \label{Eq_1_24}
\end{align}
\begin{align}  \nonumber
	s^1 & =  \frac{\hbar}{2} \sin \theta \cos \varphi = \frac{\hbar}{2} n^1  \\  \nonumber
	s^2 & =  \frac{\hbar}{2} \sin \theta \sin \varphi  = \frac{\hbar}{2} n^2  \\
	s^3 & =  \frac{\hbar}{2} \cos \theta = \frac{\hbar}{2} n^3  \label{Eq_1_25}
\end{align}
\noindent These velocities are the zitter part of the decomposition of $u^\mu(\tau),~ \mu=1,2,3$, in Eq.~\eqref{Eq_1_11} 
since the analysis here is with respect to the rest-frame $X_c$ where the global velocity is zero.
The so-called negative energy solutions, which correspond to the coefficients $A_3$ and $A_4$ in Eq.~\eqref{Eq_1_21}, are essential here; 
if they are suppressed by setting $A_3 = A_4 = 0$ in the expressions in Appendix C of {\color{Green}Beck (2023)}, 
then the zitter motion is eliminated because the velocities $u^j$ are then all zero, giving the rest-frame velocity expected for a particle without spin. 
As expected, the $s^j$ are the projections of the spin vector $\svec = \frac{\hbar}{2} \nvec$ onto the axes defined by the orthonormal triad of 3-vectors $\{\evec_1, \evec_2, \evec_3 \}$. 

It is readily shown from Eq.~\eqref{Eq_1_24} that $\uvec \cdot \nvec = 0$ so that the electron's zitter motion relative to the rest-frame $X_c$ lies in a plane orthogonal to the spin direction $\nvec$. 
Furthermore, $\uvec \cdot \uvec = c^2$ so the electron moves at the speed of light, and $\dot{\uvec} \cdot \dot{\uvec} = c^2  \omega_0^2$, which justify the constraints 
$C1: u^\mu u_\mu = 0$ and $C2: \dot{u}^\mu \dot{u}_\mu = - c^2  \omega_0^2$ that Beck imposed on his neoclassical model ({\color{Green}Beck (2023)}). 
It is shown in Appendix A that $C1$ implies that $\rvec \cdot \rvec$ is constant. 
Thus, the zitter motion is a circle about $C$, which clearly has circular frequency $\omega_0$ 
and therefore radius $r_0 = c / \omega_0 = \hbar / (2mc)$. 

This circular zitter motion produces a spin $\svec$ of magnitude $\hbar/2$ in the direction $\nvec$ orthogonal to the plane of the motion. 
Furthermore, if the electron is at space-time point $x$, Eq.~\eqref{Eq_1_14} shows that its angular position in this circular motion is 
$\omega_0 \tau(x) =  2 \theta(x)$, using Eq.~\eqref{Eq_1_10}. 
Thus, as noted in {\color{Green}Hestenes (1993)}, the electron's angular position in its zitter motion is twice its wave function phase $\theta(x)$; 
for example, if this phase increases by $\pi$, then the electron will complete a full cycle around its circular zitter path. 

This analysis confirms the nature of Schr\"{o}dinger's zitterbewegung using Dirac's equation that has been suggested by others 
based on extending classical models of the electron to include spin. 
It also confirms that an electron's spin is due to its zitter motion, which is an inherent and perpetual contribution to its path. 
Relative to a reference frame for which the global motion of the electron is not zero, the free motion will be a helix whose axis is a straight line.

\subsection{Uncertainty in Electron's Position and Velocity}

In quantum mechanics textbooks, $\psi^*(x)  \psi(x)$ is usually postulated as giving the PDF (probability density function) $\rho(x)$ of the electron's space-time position. 
However, it is seen here that it gives $\frac{1}{c} u^0(\tau(x)) = \dot{t}(\tau(x))$ and has oscillatory terms with the ultra-high zitter frequency. 
Actually, for a free electron, the PDF is usually taken as a uniform distribution since that satisfies the continuity equation, which is nothing but the stationary case 
(with respect to $\tau$) of Liouville's equation for the proper-time propagation of uncertainty arising from uncertain initial conditions in a dynamic system ({\color{Green}Beck (2023)}). 
In an em-field, the evolution of the PDF $\rho(x)$ could be accounted for by a factor $\rho(x)^{1/2}$ multiplying the wave function as in {\color{Green}Hestenes (1993, 2010)}.
In the free electron case, since the stationary solution of Liouville's equation with respect to proper time is constant, $\rho$ can be absorbed into the constant $A$ in Eq.~\eqref{Eq_1_2}.

Notice that 8 real constants specifying the spinor $A$ in the wave function in Eq.~\eqref{Eq_1_15} must be chosen 
and they then give the 8 initial conditions $\dot{x}^\mu(0) = u^{\mu}(0) = \widetilde{u}^\mu(0) = \bar{A} \hat{u}^{\mu} A$ and 
$\ddot{x}^\mu(0) =\dot{u}^\mu (0) = \widetilde{a}^\mu =  \bar{A} \hat{a}^{\mu} A$ in Eq.~\eqref{Eq_1_9}. 
The afore-mentioned normalization of spinor $A$ places one constraint on the choice of the 8 real constants.
These values are not sufficient, however, to specify the electron's worldline $x^{\mu}(\tau)$ ($\mu = 0,1,2,3$), 
which requires an additional 8 real constants to be specified, namely, $y^\mu(0)$ (which gives $x^\mu(0) = y^\mu(0) - \dot{u}^\mu(0) / \omega_0^2$, 
as seen from Eq.~\eqref{Eq_1_13}), and $\dot{y}^\mu(0) = \pi^\mu / m$, where, for a free electron, the $\pi^\mu$ are constant parameters. 
Of course, for a classical particle without spin, these are the only initial conditions that are needed to specify its motion.
Thus, even if the wave function is completely specified, there is a residual uncertainty in the electron's space-time path 
unless the appropriate initial conditions for its global motion are also specified, so, in that sense, the wave function is incomplete. 
Altogether, then, there are 16 real constants required to specify the two worldlines $x(\tau)$ and $y(\tau)$ for the electron and its inertia (or spin) center. 

The zitter (spin) motion, $z^\mu(\tau)$, cannot be observed directly because, moving at the speed of light, it is much too fast, and so the spatial location $\xvec(\tau)$ of the electron 
cannot be known more accurately than a position vector error of magnitude of the zitter motion radius $r_0 = \hbar/(2mc)$. 
Therefore, precise information about $z^\mu(0)$ and $\dot{z}^\mu(0)$ will never be available, implying that these initial conditions for the zitter motion are essentially hidden variables. 
From Eq.~\eqref{Eq_1_13}, $z^\mu(0)$ and $\dot{z}^\mu(0)$ can be expressed in terms of $\dot{u}^\mu (0) = \widetilde{a}^\mu= \bar{A} \hat{a}^{\mu} A$ and 
$u^\mu (0) = \bar{A} \hat{u}^{\mu} A$, respectively. 
As a consequence, the wave function's initial value $\phi(0) = \psi(0) = A$ can not be known precisely since, if it was, then $z^\mu(0)$ and $\dot{z}^\mu(0)$ would be known accurately.

\section{Equivalence for Free Electrons of the Dirac Equation and the Classical Dirac Particle Models}

It is readily verified from Eq.~\eqref{Eq_1_15} that $\phi(\tau)$ satisfies the equation  
\begin{equation} \label{Eq_1_26}
	 i \hbar \dot{\phi} = \hat{H} \phi
\end{equation}
This evolution equation is satisfied by the ``classical spinor'' in the electron model in {\color{Green}Barut and Zanghi (1984)}, 
which implies that for a free electron, the Dirac wave function written as a function of proper time is a Barut-Zanghi spinor. 
Beck has proved the converse: given a Barut-Zanghi spinor $\phi(\tau)$, the spinor $\psi(x) = \phi (\tau(x))$, 
where $\tau(x)$ is defined in Eq.~\eqref{Eq_1_10}, satisfies Dirac's wave equation and so is a Dirac wave function ({\color{Green}Beck (2023)}). 
This is obviously true here from reversing the procedure to get Eq.~\eqref{Eq_1_15}, that is, $\phi(\tau)$ in this equation is the spinor solution of Eq.~\eqref{Eq_1_26}, 
and so from Eq.~\eqref{Eq_1_10}, $\phi(\tau(x)) = \psi(x)$, the solution of Dirac`s equation given in Eq.~\eqref{Eq_1_2}.

This result shows for the first time that, for a free electron at least, the Barut-Zanghi classical model of the electron is equivalent to Dirac's quantum mechanics model. 
In particular, their modeling of free electron spin and zitterbewegung must be equivalent. 

The free electron models in {\color{Green}Hestenes (2010)} and in {\color{Green}Barut and Zanghi (1984)} are also equivalent, as noted in 
{\color{Green}Rodriguez et al. (1993)}, so the same equivalence with Dirac's wave equation also applies to Hestenes zitter (light-like) model. 
Both of these models use a mix of ``classical spinors'' and classical dynamic variables in their Lagrangian functions. 
The difference is that Hestenes uses his Space-time Algebra, a Clifford algebra, to express and analyze the Lagrangian function for his zitter model. 
Note that when there is no em-field, the Pauli dipole term that Hestenes adds to his Lagrangian plays no role in the dynamics. 

Barut and Cruz also examine the addition of this dipole term to the Barut-Zanghi Lagrangian function with the goal 
of empirically producing the ``anomalous'' magnetic moment that is observed experimentally, which corresponds to a gyromagnetic ratio with 
$g \approx 2(1 + \alpha/(2\pi))$ where $\alpha$ is the usual fine structure constant ({\color{Green}Barut and Cruz (1993)}). 
It is important to note, however, that Dirac's equation does not have an explicit dipole term since it uses only minimal coupling to the em-field but 
it still gives rise to magnetic and electric dipole moments, albeit with a gyromagnetic ratio corresponding to exactly $g=2$. 

From Eq.~\eqref{Eq_1_12} and Eq.~\eqref{Eq_1_13}, we see that 
\begin{align}\nonumber
	\ddot{x}^\mu (\tau) & = \dot{u}^\mu (\tau) = - \omega_0^2 z^\mu  (\tau) = - \omega_0^2 [ x^{\mu}(\tau) - y^{\mu}(\tau) ]  \  \\ \label{Eq_1_27}
	\ddot{y}^\mu(\tau) & = 0
\end{align}
\noindent giving two coupled second-order equations of motion describing the free electron that are derived from Dirac`s equation. 
The equations of motion in  Eq.~\eqref{Eq_1_27} are a special case of the neoclassical relativistic mechanics model in {\color{Green}Beck (2023)} 
where Newton's second law in proper time gives the global acceleration $\ddot{y}^\mu(\tau)$, consistent with the inertia being located at the inertia center $C$, 
the point whose spatial position is given by $\yvec(\tau)$ relative to the inertial reference frame $X_o$. His fundamental equations of motion are 
\begin{align}\nonumber
	\ddot{x}^{\mu} & = - \omega_0^2 \left( x^{\mu} - y^{\mu} \right) \  \\ \label{Eq_1_28}
	\ddot{y}^{\mu} & = \frac{1}{m} f^{\mu} \equiv \frac{q}{m} F^{\mu \nu} (x) \dot{x}_{\nu}
\end{align}
\noindent where $F(x)$ is the electromagnetic field tensor at the electron's space-time coordinates $x = (ct,\xvec)$, $\xvec$ being the spatial position of the charge. 
Thus, for the free electron, Dirac's wave equation implies Beck's neoclassical model. 

By differentiating the first of Eq.~\eqref{Eq_1_28} twice with respect to $\tau$, we can readily derive a fourth-order equation for each of the four components of the electron's space-time path $x^{\mu}(\tau)$: 
\begin{equation}\label{Eq_1_28b}
	\ddddot{x\hspace{0pt}}^{\mu} + \omega_0^2 \ddot{x}^{\mu} - \frac{q \omega_0^2}{m} F^{\mu\nu} (x) \dot{x}_{\nu} = 0
\end{equation}
\noindent 	For complete specification of the solution, these equations of motion require the 16 initial conditions: $x(0)$, $\dot{x}(0)$, $\ddot{x}(0)$, $\dddot{x\hspace{0pt}}(0)$.
These equations of motion correspond to the neoclassical Lagrangian function:
\begin{equation} \label{Eq_1_29}
	L = \frac{m} {2} u^{\mu} u_{\mu} + q A^{\mu}(x) u_{\mu} - \frac{m} {2 \omega_0^2} \dot{u}^{\mu} \dot{u}_{\mu}
\end{equation}
\noindent where $A = \left( V/c, \Avec \right)$ is the 4-vector potential for the em-field. 
This Lagrangian function is just the classical relativistic one with minimal coupling to the em-field (e.g. {\color{Green}Goldstein (1959)}) 
except for the additional last term, which turns out to be the contribution from the spin kinetic energy. 
The same Lagrangian was chosen for a free electron in {\color{Green}Riewe (1972)} and also in {\color{Green}Salesi (2002)} for his classical Dirac particle model. 
We refer to these electron models that are based on this Lagrangian as neoclassical relativistic mechanics models  
because they use classical relativistic mechanics with the addition of a new spin energy term in the Lagrangian function. 

Rivas has presented a similar relativistic model of the electron that uses an inertial observer's time $t$ rather than the proper time $\tau$ ({\color{Green}Rivas (2003)}).  
He derives it from the Frenet-Serret equations for a curve in three-dimensional space. 
By examining his equations of motion for a free electron relative to the rest-frame $X_c$, which is inertial, it can be shown that they are equivalent 
to the space part of Eq.~\eqref{Eq_1_28} in Beck's neoclassical model. 
The time part in Eq.~\eqref{Eq_1_28} is not needed in the rest-frame because observer time $t$ equals proper time $\tau$ there. 
For the interaction with an em-field, both of these theories use Newton's second law for the motion of the inertia center, with the Lorentz force evaluated at the charge center. 
Therefore, the neoclassical model of Beck can be viewed as a covariant form of the relativistic model of Rivas. 

Beck introduced a physically-based spin tensor $S$ defined by ({\color{Green}Beck (2023)})
\begin{equation} \label{Eq_1_30a}
	S^{\mu\nu} = - m \left( z^{\mu}u^{\nu} - z^{\nu}u^{\mu} \right)
\end{equation}
\noindent (The negative sign is needed to get conservation of total angular momentum for a free electron, that is, $J^{\mu\nu}(\tau) = S^{\mu\nu}(\tau) + L^{\mu\nu}(\tau)$ is constant, where 
$L^{\mu\nu} = x^\mu \pi^\nu - x^\nu \pi^\mu$ is the orbital angular momentum tensor about the origin $O$).
He then proved that with this definition, the equations of motion in Eq.~\eqref{Eq_1_28} are equivalent to the equations of motion 
\begin{align}  \nonumber
	\dot{x}^{\mu} & = u^{\mu}  \\ \nonumber
	\dot{u}^{\mu} & =  \frac{4c^2}{\hbar^2} S^{\mu \nu} \pi_{\nu}  \\ \nonumber
	\dot{S}^{\mu \nu} & = \pi^{\mu} u^{\nu} - \pi^{\nu} u^{\mu}  \\ 
	\dot{\pi}^\mu & = q F^{\mu \nu} (x) u_{\nu} \label{Eq_1_30}
\end{align}
\noindent These same equations were derived in {\color{Green}Barut and Zanghi (1984)} by applying Dirac's 4-vector velocity operator 
and his spin tensor operator to their ``classical spinor'' $\phi(\tau)$ (they did not give any proof, which is given in Section 4.2 of {\color{Green}Beck (2023)}). 
Thus, the simple electron model given in Eq.~\eqref{Eq_1_28} has the same dynamics as the Barut-Zanghi model, and so for a free electron, the same dynamics as Dirac's electron model.
Not only are the equations in Eq.~\eqref{Eq_1_30} implied by Dirac's equation and operators for a free electron, they also hold for an electron in an em-field 
provided the dynamic variables $u^{\mu}, S^{\mu \nu}$ and $\pi^\mu$ are replaced by their corresponding Dirac operators. 
This can be shown by employing the proper time derivative operator introduced in {\color{Green}Beck (1942)} to derive the Heisenberg operator equations from Dirac's equation 
(e.g. see {\color{Green}Barut (1987)}).

An important energy equation is an immediate consequence of these equations of motion:
\begin{equation}\label{Eq_1_31}
	E\dot{t} - \Pvec \cdot \dot{\xvec} = u^\mu \pi_\mu = mc^2 
\end{equation}
Firstly, the inner product is constant along an electron's worldline since $\dot{u}^{\mu} \pi_\mu = 0$ and $\dot{\pi}^{\mu} u_\mu = 0$ as a result of the spin and field tensors 
in Eq.~\eqref{Eq_1_30} being anti-symmetric. Secondly, taking a reference frame where $\Pvec = \mathbf{0}$, even if it is only instantaneously, we have $E=mc^2$ and $\dot{t} = 1$, 
showing that this constant Lorentz scalar is $mc^2$. 
Dirac's wave equation in Eq.~\eqref{Eq_1_1} can be viewed as an operator form of this equation.

\subsection{Neoclassical and Dirac's Energy Equations}

As noted above, for a free electron, the use of spinors in Dirac's wave equation and in the Barut-Zanghi evolution equation in Eq.~\eqref{Eq_1_26} is an alternative way 
to describe the same dynamics of an electron as in the various classical Dirac particle models that use only classical dynamic variables. 
This is not the case, however, for an electron in an em-field. This conclusion is based on the following argument. 

Beck derived an energy equation based on Eq.~\eqref{Eq_1_31} that extends the usual ``on-shell'' relativistic energy-momentum equation 
by an additional term arising from the electron's magnetic and electric dipole energies: 
\begin{equation} \label{Eq_1_34}
	\frac{1}{m} \pi^\mu \pi_\mu = \frac{1}{mc^2} E^2 - \frac{1}{m} \Pvec ^2 = mc^2 + \Phi
\end{equation} 
\noindent where 
\begin{equation} \label{Eq_1_35}
	\Phi \equiv - \pi^\mu \dot{z}_\mu = f^\mu z_\mu = - \frac{q}{2m} S^{\mu \nu} F_{\mu \nu} = - \frac{q}{m} \left( \Bvec \cdot \svec + \frac{1}{c} \Evec \cdot \dvec \right) 
\end{equation}
\noindent using his derived equation $\pi^\mu z_\mu = 0$ ({\color{Green}Beck (2023)}). 
The orthogonal 3-vectors $\svec = \zvec \times m \uvec$ and $\dvec = m ( u^0 \zvec - z^0 \uvec )$ are defined in 
the usual way by the space and time components of the spin tensor $S^{\mu \nu}$ defined by Eq.~\eqref{Eq_1_30a}. 
The physics behind these two vectors is more transparent when they are expressed relative to the rest-frame $X_c$ as $\svec = \rvec \times m \dot{\rvec}$ and $\dvec = mc \rvec$. 
Notice that the sum of the dipole energies is $\Phi(\tau) = \frac{1}{2} P^{\mu\nu}(\tau) F_{\mu\nu}(x(\tau))$, where $P^{\mu\nu} = - \frac{q}{m} S^{\mu\nu}$ is the dipole moment tensor, 
explicitly showing the dipole coupling to the em-field through the work done by the field on the dipole current. 

This energy equation is identical to Dirac's one that he derives in {\color{Green} Dirac (1928)} by "squaring'' his wave equation, except that his term for the dipole energies is twice as large:
\begin{equation} \label{Eq_1_32}
	\frac{1}{m} \hat{\pi}^\mu \hat{\pi}_\mu = mc^2 + \hat{\Phi}
\end{equation}
\noindent where the kinematic momentum operator is $\hat{\pi}_\mu = i \hbar \del_\mu- qA_\mu$ with $A_\mu$ the covariant components of the 4-vector em-potential, 
and $\hat{\Phi}$ is the dipole operator corresponding to the interaction energy from the electric and magnetic dipoles set up by the spin motion of the electron
\begin{equation} \label{Eq_1_33}
	\hat{\Phi}(x) = - \frac{q}{m}  \hat{S}^{\mu \nu} F_{\mu \nu}(x) = - \frac{2q}{m} \left( \Bvec \cdot \hat{\svec} + \frac{1}{c} \Evec \cdot \hat{\dvec} \right) 
\end{equation}
\noindent Here, the 3-vector operators $\hat{\svec}$ and $\hat{\dvec}$ are defined in the usual way from the space and time components of the spin tensor operator $\hat{S}^{\mu \nu}$. 
 
The factor of 2 difference between $\Phi(\tau(x))$ defined in Eq.~\eqref{Eq_1_35} and $\widetilde{\Phi}(x) = \bar{\psi}(x) \hat{\Phi}(x) \psi(x)$ from Eq.~\eqref{Eq_1_33} 
shows that for an electron in an em-field, the neoclassical and Dirac electron models are not equivalent, unlike the free electron case. 
Dirac's analysis gives no insight into the physics of the source of the dipole moment energies, which come from mathematical properties of his operators. 
On the other hand, the derivation of the neoclassical energy equation in Eq.~\eqref{Eq_1_34} shows that $\Phi$ is related to the rotational kinetic energy of the zitter motion 
and comes from the work done by the Lorentz force from the external em-field acting on the zitter component of the electron's motion to create the magnetic and electric dipole energies.
The missing factor of exactly 2 in Beck's neoclassical model is intriguing because it suggests that his model is partly capturing the physics of the dipole energy sources but not all. 
It is as if the Lorentz force acting on the zitter motion component is twice as effective as this force acting on the global motion component of the inertia (spin) center.

\subsection{Gordon Decomposition for a Free Electron}

The spin tensor $\widetilde{S}^{\mu\nu} (x)$ in Eq.~\eqref{Eq_1_9b} can be directly evaluated by substituting Dirac's wave function $\psi(x)$ from Eq.~\eqref{Eq_1_2} to get 
\begin{equation} \label{Eq_1_40b}
	\widetilde{S}^{\mu\nu}(x) = \widetilde{\Sigma}^{\mu\nu} + [\widetilde{S}^{\mu\nu}(0) - \widetilde{\Sigma}^{\mu\nu}] \cos 2 \theta(x) + \frac{1}{\omega_0} D^{\mu\nu} \sin 2 \theta(x)
\end{equation}
\noindent where $\widetilde{S}^{\mu\nu}(0) = \bar{A} \hat{S}^{\mu\nu} A$ (since $A=\psi(0)$), $D^{\mu\nu} \equiv \bar{A} \frac{i}{\hbar} [\hat{H}, \hat{S}^{\mu\nu}] A$ 
and we define the constant anti-symmetric tensor $\widetilde{\Sigma}^{\mu\nu} = \frac{1}{2} \bar{A} [ \hat{S}^{\mu\nu} + \hat{H} \hat{S}^{\mu\nu} \hat{H} / (mc^2)^2] A$.

As an alternative derivation, we can use the proved equivalence of Dirac's equation and the classical Dirac particle models of the free electron. 
Either by differentiating $\dot{S}^{\mu\nu}$ in Eq.~\eqref{Eq_1_30} and using Eq.~\eqref{Eq_1_30a}, or from Eq. (13) in {\color{Green}Barut and Zanghi (1984)} 
for $S^{\mu\nu} (\tau) = \bar{\phi}(\tau)~\hat{S}^{\mu\nu} \phi(\tau)$, we have 
\begin{equation} \label{Eq_1_41}
	\ddot{S}^{\mu\nu} + \omega_0^2 S^{\mu\nu} = \omega_0^2 \Sigma^{\mu\nu}  
\end{equation}
\noindent where $\Sigma^{\mu\nu} = - m \left( z^{\mu}\dot{z}^{\nu} - z^{\nu}\dot{z}^{\mu} \right)$, which is the constant angular momentum tensor of the electron's zitter motion about its spin center $C$. 
(Barut and Zanghi define $S^{\mu\nu}$ and $\Sigma^{\mu\nu}$ without the negative signs at the front but this implies that $\dot{S}^{\mu\nu}$ 
does not agree with their Eq. (5) and so does not give conservation of total angular momentum for a free electron, despite the statement  that it does after their Eq. (12)).
The solution of this equation is
\begin{equation} \label{Eq_1_42}
	S^{\mu\nu}(\tau) = \Sigma^{\mu\nu} + [S^{\mu\nu}(0) - \Sigma^{\mu\nu}] \cos  \omega_0 \tau + \frac{1}{\omega_0} \dot{S}^{\mu\nu}(0) \sin  \omega_0 \tau 
\end{equation}
\noindent As before, $S^{\mu\nu}(\tau(x))$ can be interpreted as $\widetilde{S}^{\mu\nu}(x)$. 
If we replace $\omega_0 \tau (x)$ in Eq.~\eqref{Eq_1_42} by $2 \theta (x)$ using Eq.~\eqref{Eq_1_10}, and note that $\tau(x) = 0$ at $x=0$, we get the same expression for 
$\widetilde{S}^{\mu\nu} (x)$ as in Eq.~\eqref{Eq_1_40b}. 
Notice that equating the two expressions for $S^{\mu\nu}(\tau(x))$ from Eq.~\eqref{Eq_1_40b} and Eq.~\eqref{Eq_1_42} implies that 
$\widetilde{\Sigma}^{\mu\nu} = \Sigma^{\mu\nu}$, $\widetilde{S}^{\mu\nu}(0) = S^{\mu\nu}(0)$ and $D^{\mu\nu} = \dot{S}^{\mu\nu}(0)$ because the functions 
$\{1, \cos \omega_0 \tau, \sin  \omega_0 \tau\}$ are linearly independent. We note that $\dot{S}^{\mu\nu}(0) = D^{\mu\nu} \equiv \bar{A} \frac{i}{\hbar} [\hat{H}, \hat{S}^{\mu\nu}] A$ 
is consistent with the operator for a derivative implied by Eq.~\eqref{Eq_1_26} and given in Eq. (4.4) in {\color{Green}Beck (2023)}. 

Differentiating the expression for $\widetilde{S}^{\mu\nu}(x)$ in Eq.~\eqref{Eq_1_40b} to get the second term in the Gordon decomposition in Eq.~\eqref{Eq_1_9b}
\begin{equation} \label{Eq_1_43}
	- \frac{1}{m} \del_\nu \widetilde{S}^{\mu\nu}(x) = \frac{2}{m \hbar} [S^{\mu\nu}(0) \pi_\nu - \Sigma^{\mu\nu}\pi_\nu] \sin  2\theta(x) - \frac{2}{m \hbar \omega_0} \dot{S}^{\mu\nu}(0) \pi_\nu \cos  2\theta(x)  
\end{equation}
\noindent As shown in {\color{Green}Beck (2023)} in Eq. (2.15), $z^\mu \pi_\mu = 0$, implying by differentiation that $\dot{z}^\mu \pi_\mu = 0$ since $\dot{\pi}_\mu = 0$ for a free electron. 
It then follows from the definition of $\Sigma^{\mu\nu}$ that $\Sigma^{\mu\nu}\pi_\nu = 0$. 
It is also shown in Beck's Eq. (3.11)(iii) that $S^{\mu\nu} \pi_\nu = - (mc)^2 z^\mu$, which implies by differentiation that $\dot{S}^{\mu\nu} \pi_\nu = - (mc)^2 \dot{z}^\mu$.
Substituting these results into Eq.~\eqref{Eq_1_43} 
\begin{equation} \label{Eq_1_44}
	\widetilde{w}^\mu(x) = - \frac{1}{m} \del_\nu \widetilde{S}^{\mu\nu}(x) = \dot{z}^\mu (0) \cos  2\theta(x) - \omega_0 z^\mu(0) \sin  2\theta(x) \equiv \dot{z}^\mu (\tau(x)) 
\end{equation}
\noindent From the definition of $z^\mu(\tau)$ in Eq.~\eqref{Eq_1_13}, $z^\mu(0) = - \dot{u}^\mu(0) / \omega_0^2 = - \widetilde{a}^\mu / \omega_0^2$ and 
$\dot{z}^\mu(0) = u^\mu(0) - \pi^\mu / m = \widetilde{u}^\mu(0) - \pi^\mu / m$, so Eq.~\eqref{Eq_1_44} is indeed equivalent to $\widetilde{w}^\mu(x)$ in Eq.~\eqref{Eq_1_9c}.

The structure of the 4-tensor for spin implies $q \widetilde{w}^\mu (x) = - \frac{q}{m} \del_\nu \widetilde{S}^{\mu\nu} (x)$ can be expressed in terms of its time and space parts as 
\begin{equation} \label{Eq_1_47}
	q \widetilde{w}(x) = (q \widetilde{w}^0(x), q \widetilde{\wvec} (x)) = \frac{q}{m} \big( - \nabla \cdot \dvec(x), \frac{1}{c} \frac{\del}{\del t} \dvec(x) + \nabla \times \svec(x) \big)
\end{equation}
\noindent The polarization vector is $\frac{1}{mc} \dvec = \dot{t} \zvec - t_z \uvec$, which is just the zitter radius vector $\rvec$ in the rest-frame $X_c$ 
where $\yvec=0$ and so it is the electron's spatial position $\xvec$ relative to $X_c$.  
The spatial part $q \widetilde{\wvec}(x)$ is the sum of polarization and magnetization currents (e.g. {\color{Green}Baym (1981)}). 
Therefore, these currents are a consequence of the zitter motion of the electron. 
It is a curious result that the magnetization current $\frac{q}{m} \nabla \times \svec$ is zero in the rest-frame for a free electron since the spin vector $\svec = \bar{\psi}(x) \hat{\svec} \psi(x)$ is constant there. 
However, the circular zitter motion in the rest-frame is still present, of course, as are the magnetic and electric dipole moment 3-vectors, 
\mbox{\boldmath $\mu$} $= \frac{q}{m} \svec$ and \mbox{\boldmath $\epsilon$} $= \frac{q}{mc} \dvec$ ($= q \rvec$), respectively.
The zitter motion then corresponds entirely to the polarization current, which is from Eq.~\eqref{Eq_1_47}, $q \wvec =  \frac{q}{mc} \frac{\del}{\del t} \dvec = q\dot{\rvec}$ since $t = \tau$ in the rest-frame. 
The spin axis has to precess for the magnetization current to be non-zero, as it does for a free electron when viewed by an observer moving relative to the electron's rest-frame.

The neoclassical model decomposition of the electron's space-time position $x = y + z$ provides an integrated form for the Gordon decomposition 
\begin{equation} \label{Eq_1_45}
	x^\mu = y^\mu - \frac{1}{m^2 c^2} S^{\mu\nu} \pi_\nu 
\end{equation}
\noindent where the time and space parts of the last term are
\begin{equation} \label{Eq_1_46}
	z = (z^0, \zvec) = \frac{1}{m^2 c^2} (\dvec \cdot \Pvec, \frac{E}{c} \dvec + \svec \times \Pvec)
\end{equation}
\noindent This result holds in an em-field and differentiation of Eq.~\eqref{Eq_1_45} with respect to proper time gives a neoclassical Gordon decomposition of Dirac's current $u^\mu$ for this case.

\section{Concluding Remarks}

In this work, we show for the first time that the classical particle model of the Dirac electron in {\color{Green}Barut and Zanghi (1984)} that uses  
``classical'' spinors and classical dynamic variables of a proper time, is equivalent to Dirac's equation in the case of a free electron, that is, 
every Barut-Zanghi spinor solution is a Dirac wave function solution, and conversely. 
It has already been shown in {\color{Green}Beck (2023)} that there is an equivalence between the equations of motion in classical dynamic variables 
that are derived by applying Dirac's velocity and spin operators to the Barut-Zanghi spinor, and the equations of motion of Beck's neoclassical relativistic mechanics model of the electron. 
In addition, it is noted in this work that the classical particle models in {\color{Green}Hestenes (2010)} and {\color{Green}Barut and Zanghi (1984)} that both use spinors to express the dynamics 
of an electron are equivalent, and that the classical spin models in {\color{Green}Beck (2023)}, {\color{Green}Rivas (2003)} and {\color{Green}Salesi (2002)} are also equivalent. 
Therefore, all these classical models describe the same dynamics for the spinning electron, and, for the free electron, they all agree with the dynamics expressed by Dirac's wave equation. 

The free electron paths extracted from solutions of Dirac's equation in this work confirm an underlying reality for the nature of the electron's spin and the phenomenon of zitterbewegung that has 
previously been revealed by the afore-mentioned classical spin models, but a reality that has not been widely accepted, 
no doubt because the equivalence of these classical models with Dirac's equation was doubted. These electron paths reveal the following:

\begin{itemize} 
	\item[--] The electron's spin, which is known to not be a simple rotation of the electron, is a perpetual local motion 
	about a globally moving spin center that is inherent in the electron's space-time trajectory as a ``point'' particle.
	For a free electron, when viewed from a rest-frame fixed at the spin center, this spin motion is circular in a plane orthogonal to a constant spin axis and it 
	occurs at the speed of light with the ultra-high zitterbewegung circular frequency. However, the observed global speed of the spin center is always sub-luminal 
	with the difference between the speeds of the electron and spin center being accounted for by the local spin motion. 
	The electron's inertia is effective at its spin center, whose worldline is controlled by Newton's second law in proper time, the time of a clock at the spin center,
	and so it is appropriately called the inertia center (or center of mass, although it is not technically one).

	\item[--] The  mysterious phenomenon of zitterbewegung that was revealed by Schr\"{o}dinger using a Heisenberg operator analysis of Dirac's wave equation for the free electron, 
		is shown to be a manifestation of its local spin motion, which may also be viewed as the mechanism for de Broglie's ``internal clock'' of the electron. 
		Therefore, this spin motion is appropriately called the ``zitter motion'' by David Hestenes.		
	\item[--] The so-called polarization and magnetization currents in the Gordon decomposition of Dirac's current are due to the zitter motion, which can also be viewed as giving a dipole current. 
		
	\item[--] The phase $\theta(x)$ of the Dirac wave function in Eq.~\eqref{Eq_1_3} is 1/2 of the angular position of the electron in its zitter loop.

\end{itemize}

Furthermore, since Dirac's equation for a free electron implies the equations of motion of the neoclassical model presented in {\color{Green}Beck (2023)}, 
his previous conclusions also follow from Dirac's equation, such as 

\begin{itemize} 	
	\item[--] An apparent plane-wave characteristic of the electron's motion, consistent with de Broglie's wave theory, is just the local spin motion viewed 
	through the Lorentz transformation by an observer fixed with respect to another reference frame. 
	
	\item[--] The dependence of the electron's kinetic energy on an observer's reference frame is accounted for by its storage as the spin rotational energy $\hbar \omega /2 = \gamma m c^2$ 
	(like a flywheel) with an observed angular speed $\omega = \gamma \omega_0$ that depends on the velocity of the electron's inertia center relative to the observer's reference frame. 
	
	\item[--] Dirac's spin tensor operator represents the angular momentum of the electron's total velocity about its spin (or inertia) center. 
	For the conservation of angular momentum of the free electron about its spin center, this operator needs a change of sign from the usual definition.
	 
	\end{itemize}

The theory presented here for extracting the electron's worldline from Dirac's equation is in the spirit of a relativistic Bohmian mechanics in that it determines the electron's motion, 
including its spin, by applying Dirac's velocity operator to his free-electron wave function to elucidate the nature of spin and zitterbewegung.
Whether this approach to revealing electron paths can be extended to explain the electron's behavior in cases where it interacts with an em-field remains to be determined. 
This would be especially interesting to do for an electron moving in a magnetic field because it might explain why there is a factor of 2 difference 
between the Dirac and neoclassical terms for the magnetic dipole energy in their respective energy equations. 

One obvious challenge to extending this work for an electron in an em-field is that, unlike the free electron case, there is no analytical solution $\psi(x)$ of Dirac's equation, in general, 
although there is for a uniform magnetic field, suggesting a fruitful direction for future research.
When an analytical solution is available, Dirac's velocity and spin operators can be applied to the wave function to evaluate the proper velocity $\widetilde{u}^\mu(x)$ for the electron 
at space-time point $x$, along with the spin tensor $\widetilde{S}^{\mu\nu}(x)$. 
The Gordon decomposition in Eq.~\eqref{Eq_1_9b} should then give the inertia center velocity $\widetilde{v}^\mu(x) = \widetilde{\dot{y}}^\mu(x)$ 
and the zitter (spin) velocity $\widetilde{w}^\mu(x) = \widetilde{\dot{z}}^\mu(x)$. 

Another challenge in this case is that the relation between the Dirac wave function phase $\theta(x)$ and the proper time $\tau(x)$ in Eq.~\eqref{Eq_1_10} 
must be expressed in incremental form and integrated along the electron's worldline to get $\tau(x)$, so they must be established simultaneously. 
In Beck's classical Dirac particle model covered in this paper, this worldline, which gives the charge position in proper time, is governed by a fourth-order ordinary differential equation 
that can be expressed as two coupled second-order ordinary differential equations that also includes the worldline for the electron's inertia (or spin) center.
It remains to be seen if that is also true for Dirac's equation when an em-field is present. 

It also remains to be determined if the approach can be extended to $N$ interacting electrons. 
One challenge in this case is that proper time is specific to an electron and so there will be $N$ proper times, each of which must be related to an inertial observer's time. 
Of course, for this case, Dirac's single-electron equation is usually replaced by a quantum field theory interpretation of it.

\appendix
\section{}

Consider the zitter motion relative to the rest-frame $X_c$ with the electron moving at the speed of light, so $\uvec \cdot \uvec = c^2$, which gives by differentiation $\dot{\uvec} \cdot \uvec = 0$. 
From Eq.~\eqref{Eq_1_14}, $\rvec = - \ddot{\rvec} / \omega_0^2 = - \dot{\uvec} / \omega_0^2$, and we see that $\rvec \cdot \uvec = 0$. 
Since $\uvec = \dot{\rvec}$, this implies that $\rvec \cdot \rvec$ is constant. The zitter motion is therefore a circle about $C$, which, 
from Eq.~\eqref{Eq_1_14}, has a circular frequency $\omega_0$ and therefore radius $r_0 = c / \omega_0 = \hbar / (2mc)$.

\section*{References}

\noindent A.O. Barut (1987), Electron as a radiating and spinning dynamical system and discrete internal quantum systems, Phys. Scripta, 35, 229-232. \\

\noindent A.O. Barut and M.G. Cruz (1993), Classical relativistic spinning particle with anomalous magnetic moment: the precession of spin, J. Phys. A: Math. Gen. 26, 6499-6506. \\

\noindent A.O. Barut and N. Zanghi (1984), Classical model of the Dirac electron, Phys. Rev. Lett. 52, 2009-2012. \\

\noindent A.O. Barut and W. Thacker (1985), Covariant generalization of the Zitterbewegung of the electron and its SO(4,2) and SO(3,2) internal algebras, Phys. Rev. D 31, 1386-1392. \\

\noindent G. Baym (1981), Lectures in Quantum Mechanics, 9th printing, W.A. Benjamin Inc., Reading, MA, USA. \\

\noindent G. Beck (1942), Rev. Faculdade de Ciencias de Coimbra, 10, 66. \\

\noindent J.L. Beck (2023), Neo-classical relativistic mechanics theory for electrons that exhibits spin, zitterbewegung, dipole moments, 
wavefunctions and Dirac's wave equation, Foundation Phys. 53:57, 1-39. \\

\noindent J.D. Bjorken and S.D. Drell (1964), Relativistic Quantum Mechanics, McGraw-Hill, New York, USA. \\

\noindent D. Bohm and B.J. Hiley (1993), The Undivided Universe, Routledge, London, UK \\

\noindent P.A.M. Dirac (1928), The quantum theory of the electron, Proc. Roy. Soc. London A117, 610. \\

\noindent H. Goldstein (1959), Classical Mechanics, Addison-Wesley Publishing, Reading, MA, USA. \\

\noindent D. Hestenes (1993), Zitterbewegung modeling, Foundation Phys. 23, 365-387. \\

\noindent D. Hestenes (2010), Zitterbewegung in quantum mechanics, Foundation Phys. 40, 1-54. \\

\noindent K. Huang (1952), On the zitterbewegung of the free Dirac electron, Am. J. Phys. 20, 479-484. \\

\noindent A. Peletminskii and S. Peletminskii (2005), Lagrangian and Hamiltonian formalisms for relativistic dynamics of a charged particle with dipole moment, Eur. Phys. J. C 42, 505-517. \\

\noindent F. Riewe (1972), Relativistic classical spinning-particle mechanics, Il Nuovo Cimento 8B, 271-277. \\


\noindent M. Rivas (2003), The dynamical equation of the spinning electron, J. Phys. A: Math. Gen. 36, 4703-4716. \\

\noindent M. Rivas (2015), The center of mass and the center of charge of the electron, J. Phys.: Conf Set 615 012017. \\

\noindent W.A. Rodrigues, J.J Vaz, E. Recami and G. Salesi (1993), About zitterbewegung and electron structure, Phys. Lett. B. 318, 623-628. \\

\noindent G. Salesi (2002), Non-Newtonian mechanics, Int. J. Mod. Phys. A 17, 347-374. \\

\noindent E. Schr\"{o}dinger (1930), \"{U}ber die Kr\"{a}fetfreie Bewegung in der relativistischen Quantenmechanik, Sitzungsber. Preuss. Akad. Wiss. Physik-math. K1, 24, 418-428. \\

\end{document}